\documentstyle[11pt,twoside,epsfig,jltp]{article}

\title{Unusual Charge Localization in Zn-doped and Heavily Underdoped
YBa$_{2}$Cu$_{3}$O$_{7-\delta}$ \\ at Low Temperatures}
\author{Kouji Segawa and Yoichi Ando \address{
Central Research Institute of
Electric Power Industry, Tokyo 201-8511, Japan}}

\runninghead{Kouji Segawa and Yoichi Ando}{
Unusual Charge Localization in YBCO}

\begin{document}

\begin{abstract}
The in-plane normal-state resistivity of Zn-doped YBa$_{\it 2}$Cu$_{\it 3}$O$_{\it 7-\delta}$
and heavily underdoped
pure YBCO single crystals
is measured down to low temperatures under magnetic fields up to 18 T.
We found that the temperature dependence of the normal-state $\rho_{\it ab}$
does not obey $\log (1/T)$ and tends not to diverge in the low temperature limit.
The result suggests that the ``ground state'' of the normal state of YBCO is metallic.


\end{abstract}

\maketitle

\vspace{1cm}

In high-$T_c$ cuprates,
the low-temperature normal-state resistivity
is expected to reflect the electronic structure which underlies the high-$T_c$ superconductivity.
In La$_{2-x}$Sr$_x$CuO$_4$ (LSCO) system,
the measurement of the normal-state resistivity
at low temperatures was performed using a pulsed magnet \cite{Ando}.
The in-plane resistivity $\rho_{\it ab}$ of underdoped LSCO
was reported to show logarithmic divergence
in high magnetic fields in the zero-temperature limit. 
Thus, the ``ground state'' of the normal state of the underdoped LSCO appears
to be insulating.
In order to see whether the insulating ``ground state'' is
a common feature of the underdoped high-$T_c$ cuprates or not,
one needs to perform the measurement in other high-$T_c$
systems.
In the case of YBa$_2$Cu$_3$O$_{7-\delta}$ (YBCO), the normal-state $\rho_{\it ab}$
has been measured at low temperatures by using 18 T magnetic field \cite{Segawa}.
Zn substitution in excess of 2.5\%
was reported to induce an upturn in 
the temperature dependence of $\rho_{\it ab}$ at low temperatures,
and the logarithmic temperature dependence of $\rho_{\it ab}$
was observed; however,
it remains unclear whether the logarithmic divergence
continues to very low temperatures in the YBCO system,
because the 18 T magnetic field was insufficient to completely suppress superconductivity
at low temperatures in superconducting samples.
On the other hand, the superconducting fluctuation in non-superconducting samples
can easily be suppressed by 18 T field at low temperatures.
Thus the temperature dependence of the resistivity in such non-superconducting samples
can be measured down to low temperatures without significant difficulties.
If one investigates the temperature dependence of the resistivity
in the non-superconducting samples as well as the superconducting Zn-doped samples,
the behavior of $\rho_{\it ab}(T)$ in the low temperature limit
is expected to be clarified.

Here we report measurements of the low-temperature
normal-state resistivity along the CuO$_2$ planes, $\rho_{\it ab}$,
of Zn-doped YBCO and non-superconducting heavily-underdoped pure YBCO.
We found that the normal-state $\rho_{\it ab}$ is not likely to diverge
in the low-temperature limit.
The result suggests that the ``ground state'' of the normal state is metallic in YBCO.

\begin{figure}
\centerline{\psfig{file=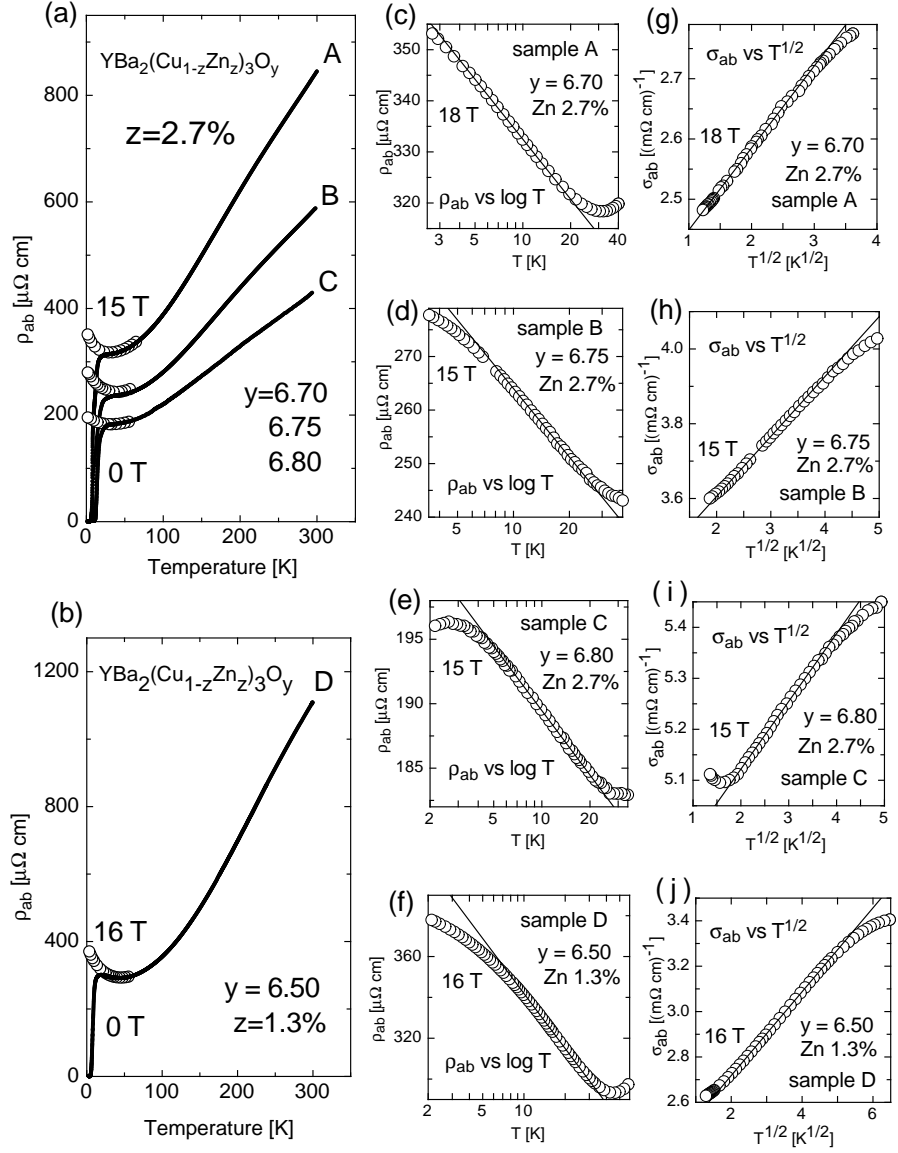,width=12.0cm}}
\caption{\small (a): $\rho_{\rm ab}(T)$ of $z$=2.7\% crystals
with different $y$ (6.70(sample A), 6.75(sample B), and 6.80(sample C))
in 0 T (solid lines) and in 15 T (open circles).
(b): $\rho_{\it ab}(T)$ in 0, 16 T
for $y$= 6.50, $z$=1.3\% (sample D).
(c)(d)(e)(f): $\log T$ plots of $\rho_{\it ab}(T)$ for samples A, B, C and D.
(g)(h)(i)(j): $\sigma_{\it ab}(T)$ vs $T^{1/2}$ plots for samples A, B, C and D.
}
\label{fig1}
\end{figure}

The single crystals of YBa$_{2}$(Cu$_{1-z}$Zn$_z$)$_3$O$_{7-\delta}$
are grown by flux method in pure
Y$_2$O$_3$ crucibles to avoid inclusion of any impurities other than Zn
\cite{Segawa}.
The oxygen content $y$ ($\equiv$ 7-$\delta$) in the crystals is
controlled by annealing in evacuated and sealed quartz tubes
at 500-650$^{\circ}$C for 1-2 days
together with sintered blocks and/or powders
and then quenched with liquid nitrogen.
The final oxygen content is confirmed by
iodometric titration with an accuracy of better than $\pm 0.01$.
The measurement of $\rho_{\it ab}$ is performed
with ac four-probe technique
under dc magnetic fields up to 18 T applied along the $c$ axis.

Figure 1(a) shows the temperature dependence of $\rho_{\it ab}$ in
0 and 15 T, for $z$=2.7\% samples with $y$=6.70 (sample A), 6.75 (sample B)
and 6.80 (sample C), and Fig.1(b) shows $\rho_{\it ab}(T)$
in 0 and 16 T for a sample with $z$=1.3\% $y$=6.50 (sample D).
The $\rho_{\it ab}(T)$ of these four samples all show
an upturn in magnetic fields at low temperatures.
Figs.1(c)(d)(e)(f) show the $\log T$ plots of $\rho_{\it ab}(T)$
for samples A, B, C and D, respectively.
The temperature dependences of $\rho_{\it ab}$ of these samples 
can be seen to be consistent with $\log T$ only in the temperature range of
5-12 K, 9-20 K, 7-15 K and 12-20 K, respectively, for samples A, B, C and D.
Below those temperature regions, $\rho_{\it ab}(T)$ starts to deviate downwardly from $\log T$.
The downward deviation is due to a different temperature dependence
of $\rho_{\it ab}$ at lower temperatures and/or
the superconducting fluctuation which is not sufficiently suppressed in 15-18 T
magnetic fields.
Figures 1(g)(h)(i)(j) show $\sigma_{\it ab}(T)$ vs $T^{1/2}$ plots for samples A, B, C and D.
The temperature dependences of $\rho_{\it ab}$
are consistent with $\sigma_{\it ab}(T) \sim a + T^{1/2}$,
in the range of 1.5-8 K (1.2-2.8 K$^{1/2}$), 6-12 K (2.4-3.5 K$^{1/2}$), 5-9 K (2.2-3.0 K$^{1/2}$)
and 8-20 K (2.8-4.5 K$^{1/2}$),
respectively, for samples A, B, C and D.
In particular, $\sigma_{\it ab}(T)$ of sample A is well fitted with
$\sim a+T^{1/2}$ down to 1.5 K.
This temperature dependence, $\sigma_{\it ab}(T) \sim a + T^{1/2}$,
is fundamentally different from that of $\rho_{\it ab}(T) \sim \log (1/T)$;
$\log(1/T)$ diverges as $T \rightarrow 0$, whereas $1/(a + T^{1/2})$ remains finite.
Thus, the result of Fig.1(g),
in which $\sigma_{\it ab}(T) \sim a + T^{1/2}$
well fits the measured data, suggest that
the in-plane resistivity in YBCO is not likely to diverge in the zero-temperature limit.

However, the resistivity might be reduced by the superconducting
fluctuation at low temperatures in those samples.
If so,
it is possible that the intrinsic
$\rho_{\it ab}(T)$ is diverging in the low temperature limit
if there were no superconducting fluctuations.
In order to avoid such uncertainty, we prepared non-superconducting
samples by reducing oxygen in pure YBCO.
Fig.2(a) shows $\rho_{\it ab}(T)$ in 0 T
for pure samples with $y$= 6.40 (sample E) and 6.38 (sample F).
Sample E is close to the superconducting region and
its resistivity value corresponds to $k_{F}l \sim 2$,
where $k_F$ is the Fermi wave number and $l$ is the mean free path.
Sample F is strongly insulating and its resistivity corresponds to $k_{F}l <1$
at low temperatures.
Both samples E and F show no superconducting transition.
However, in sample E, $\rho_{\it ab}$ shows a decrease
due to the superconducting fluctuation in 0 T at low temperatures.
The superconducting fluctuation in sample E can easily be suppressed in 18 T down to 0.2 K.

\begin{figure}
\centerline{\psfig{file=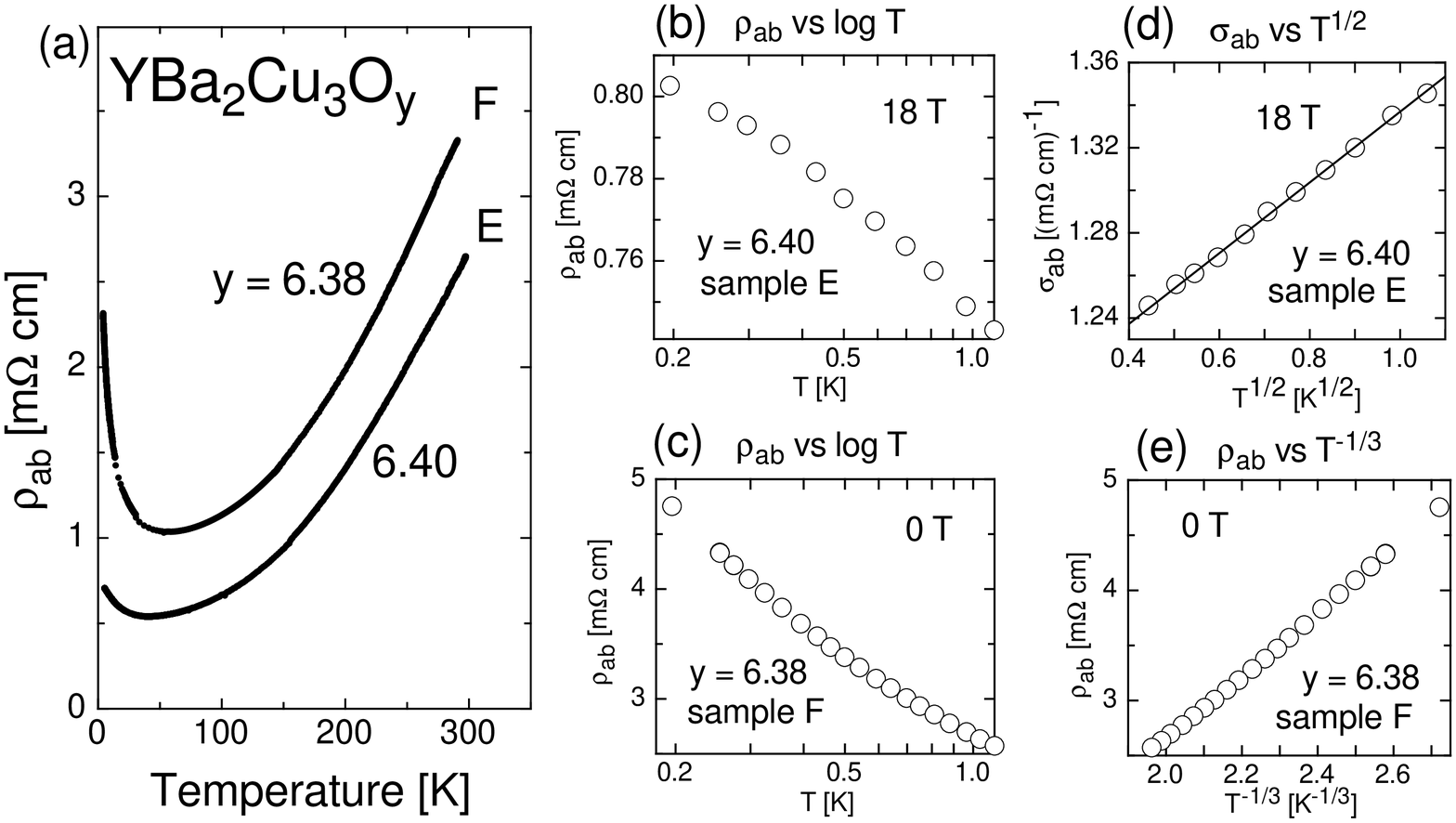,width=12.5cm}}
\caption{\small (a): $\rho_{\it ab}(T)$ in 0 T
for pure samples with $y$= 6.40(sample E), 6.38(sample F).
(b)(c): $\log T$ plots of $\rho_{\it ab}(T)$ for samples E and F.
(d): $\sigma_{\it ab}(T)$ vs $T^{1/2}$ plot for sample E.
(e): $\rho_{\it ab}(T)$ vs $T^{-1/3}$ plot for sample F.}
\label{fig2}
\end{figure}

Figure 2(b) shows the $\log T$ plot of $\rho_{\it ab}(T)$ for sample E.
This $\rho_{\it ab}(T)$ is not consistent with $\log (1/T)$.
In contrast, as Fig.2(d) shows, $\sigma_{\it ab}(T)$ is
well fitted with
$\sim a + T^{1/2}$ down to 0.2 K
(0.45 K$^{1/2}$).
We note that Lavrov {\it et al}. reported similar $T$-dependence of conductivity
in heavily underdoped YBCO\cite{Lavrov}, whose oxygen content
is between samples E and F.
However, their data were also consistent with $\rho_{\it ab}(T) \sim \log(1/T)$
as well as $\sigma_{\it ab}(T) \sim a + T^{1/2}$ at low temperatures.
In contrast to their data, our data definitely show that the increase in $\rho_{\it ab}$
with decreasing temperature is weaker than $\log T$
at low temperatures.
Thus, we may conclude that $\rho_{\it ab}$ in sample E is not diverging
in the low temperature limit.

For sample F, Fig.2(c) shows the $\log T$ plot of $\rho_{\it ab}(T)$.
It is clear in Fig.2(c) that
the resistivity of sample F
tends to diverge more strongly than $\log T$
with decreasing temperature.
We found instead that $\rho_{\it ab}(T)\sim T^{-1/3}$
is a better description of the temperature dependence of $\rho_{\it ab}$
of sample F.
Fig.2(e) shows the plot of $\rho_{\it ab}(T)$ vs $T^{-1/3}$ for sample F,
where one can find that the data lie on a straight line reasonably well
at low temperatures (right hand side of the plot).
Thus, the resistivity of sample F is likely to be diverging in the low temperature limit.
In both the superconducting Zn-doped samples and the non-superconducting
sample (but close to the superconducting composition),
$\rho_{\it ab}$ does not behave as $\log(1/T)$ at low temperatures.
Instead, the behavior of $\sigma_{\it ab}(T) \sim a + T^{1/2}$
is observed in the low temperature limit.
This result suggests that $\rho_{\it ab}$
does not diverge in the low temperature limit
and thus
the ``ground state'' of the normal state of YBCO is metallic.
Of course, not all YBCO samples are ``metallic'', for example,
$y$=6.38 sample remains insulating at low temperatures.
The resistivity of the insulating $y$=6.38 sample corresponds to $k_{F}l<1$,
whereas that of the ``metallic'' $y$=6.40 sample corresponds to $k_{F}l\sim 2$.
This means that the samples with $k_{F}l<1$ cannot be metallic
at low temperatures,
like conventional metals.

In summary, we measured the in-plane resistivity of Zn-doped YBCO
and non-superconducting heavily underdoped 
YBCO crystals down to low temperatures under magnetic fields up to 18 T.
It is found that
the temperature dependence of the normal-state $\rho_{\it ab}$
tends not to diverge
in the low temperature limit.
The result suggests that the ``ground state'' of the normal state of YBCO
is metallic.

We thank A. N. Lavrov for fruitful discussions and experimental assistance.

%

\end{document}